\documentclass[11pt]{article} 
\usepackage{appb,psfig}
\begin{document}

\title{POLARIZED PARTON DISTRIBUTIONS AND QCD SPIN TESTS AT  RHIC-BNL\thanks{Presented at The Cracow Epiphany Conference on Spin Effects in 
Particle Physics, Cracow, Poland, January 9-11, 1998.}
}

\author{J. SOFFER
\address{Centre de Physique Th\'eorique, CNRS Luminy, Case 907,\\                F-13288 Marseille Cedex 09 France} 
}
\maketitle 
\begin{abstract}
The {\it RHIC} facility at {\it BNL} will be operating soon, part of the year, as a polarized proton-proton collider. This will allow the undertaking of a vast spin physics programme, mainly by the two large detectors {\it PHENIX} and {\it STAR}. We
review some theoretical aspects of this research programme which will allow, firstly to 
improve our present knowledge on polarized quark, gluon and sea distributions in a nucleon, secondly to perform novel {\it QCD} spin tests and finally, perhaps, to 
uncover some new physics. 
\end{abstract}
  
\section{Introduction}
Considerable progress have been made over the last ten years or so, in our 
understanding of the spin structure of the nucleon. This is essentially 
due to a better determination of the polarized structure functions 
$g_1^{p,n,d}(x,Q^2)$, 
from polarized Deep-Inelastic-Scattering ({\it DIS}) on different targets 
(Hydrogen, Deuterium, Helium-3). However these fixed polarized targets 
experiments \cite{Al}, performed at {\it CERN}, {\it DESY} and {\it SLAC}, cover only a limited kinematic region, that is $0.005\leq x \leq 0.7$, with the 
corresponding average $<Q^2>$ between $2GeV^2$ and $10GeV^2$. In spite of the 
constant progress realized in the accuracy of the data, they can still be 
described, non uniquely, in terms of several sets of polarized parton 
distributions. In particular, sea quark, antiquark and gluon distributions 
remain highly ambiguous. The restricted $Q^2$ range accessible by the data makes also rather difficult, sensible tests of the $Q^2$ evolution, predicted by 
recent higher order {\it QCD} calculations. Moreover it is not possible to 
obtain a good flavor separation, to isolate the contribution of each quark to 
the nucleon spin.

Polarized hadronic collisions, which are another way to investigate this 
research field, have accomplished little progress due to the scarcity of the 
data in appropriate kinematic regions, and a low energy range, so far accessible to very few dedicated experiments. Let us recall that the highest energy for 
studying polarized $pp$ ($\bar pp$) collisions has been obtained at Fermilab by
 the {\it E704} experiment \cite{Br} with a $200GeV/c$ polarized proton 
(antiproton) beam on a fixed target, that is $\sqrt s =19.4GeV$. This situation 
will change drastically soon when the {\it RHIC} facility at {\it BNL} will 
start running, by 1999 or so, part of the time as a polarized $pp$ collider. A 
vast spin programme will be undertaken by the two large detectors {\it PHENIX} 
and {\it STAR}, which will operate at {\it RHIC} and also by the  $pp2pp$ 
experiment, dedicated to $pp$ elastic and total cross sections. Before we go on 
and explain very briefly what will be done, let us recall three key parameters, 
which will be crucial to answer some of the very challenging questions. The 
proton beam polarization $P$
will be maintained at the level of 70\%, in both {\it longitudinal} and {\it 
transverse} directions, the center-of-mass energy $\sqrt s$ will be ranging
between $100GeV$ and $500GeV$ and at its maximum value, the luminosity is 
expected to reach {\it L}=$2.10^{32}cm^{-2}sec^{-1}$.

The {\it Siberian snakes} magnets which preserve the degree of polarization in 
the {\it RHIC} main rings and the {\it spin rotators} which select the beam spin direction, are under construction thanks to a substantial financial contribution from the Japanese Institute {\it RIKEN} in collaboration with {\it BNL}. The 
high luminosity will allow very copious effective yields for different reactions ($\gamma$, jet, $W^{\pm}$ production, etc...) and therefore the measurement of 
the corresponding spin asymmetries will be made to a very good level of 
accuracy, in the kinematic regions relevant for {\it QCD} spin tests. The spin 
programme at {\it RHIC} will provide answers to fundamental questions which will be listed now in turn.

In the next section we will recall some basic definitions of the helicity 
asymmetries. Section 3 will be devoted to prompt photon production and jet 
production, which will allow the first direct determination of the gluon 
helicity distribution $\Delta G(x,Q^2)$ inside a polarized nucleon. Next we will show in section 4, how antiquark helicity distributions $\Delta \bar q(x,Q^2)$ 
can be isolated in $W^{\pm}$ production, which leads also to the {\it u-d} 
flavor separation. This has been done, rather inaccurately, in semi-inclusive 
{\it DIS}. From transversely polarized proton beams, as we will see in section 
5, it is possible to make the first measurement of the transversity 
distributions $h_1^{q,\bar q}(x,Q^2)$ in Drell-Yan lepton pair production. 
Finally, in section 6 we will indicate the relevance of the parity violating 
asymmetry in single jet production. It might provide a clean signature for 
new physics and, as an example, we will consider the possible effects of a quark-quark contact interaction.

\section{Basic definitions and helicity asymmetries}

Fundamental interactions at short distances which are explored in high energy 
hadronic collisions, involve hard scattering of quarks, antiquarks and gluons.
Let us consider the general hadronic reaction
\begin{equation}
a\, +\, b \rightarrow \, c + \, X
\end{equation}
\noindent
where $c$, in the cases we will consider below, is either a photon, a $Z$, a
$W^{\pm}$ or a single-jet. In the hard scattering kinematic region, the cross
section describing (1) reads in the {\it QCD} parton model, provided factorization
holds, as
\begin{equation}
 d\sigma (a+b \rightarrow c+X)\!\! =\!\!\! 
\sum_{ij}\!{1\over1+\delta_
{ij}}\!\!\int\!\! dx_adx_b\biggl[\!f^{(a)}_i\bigl(x_a,Q^2\bigl)
f^{(b)}_j\bigl(x_b,Q^2\bigl)\,d\hat{\sigma}^{ij}                                                                 +(i\leftrightarrow j)\!\biggl].
\end{equation}

The summation runs over all contributing parton configurations, the
$f(x,Q^2)$'s are the parton distributions, directly extracted from 
{\it DIS} for quarks and antiquarks and indirectly for gluons. Here
d${\hat \sigma}_{ij}$ is the cross section for the interaction of two partons
$i$ and $j$ which can be calculated perturbatively, some of which , at the 
lowest order, are given in ref.\cite{BRST}.
If we consider the reaction (1) with {\it {both}} initial hadrons, $a$ and 
$b$ longitudinally polarized, one useful observable is the {\it{double}}
helicity asymmetry $A_{LL}$ defined as
\begin{equation}
A_{LL} ={d\sigma_{a(+)b(+)}-d\sigma_{a(+)b(-)}\over 
d\sigma_{a(+)b(+)}+d\sigma_{a(+)b(-)}}\; ,
\end{equation}
when we assume parity conservation, i.e.
d$\sigma_{a(\lambda)b(\lambda')}$ = d$\sigma_{a(-\lambda)b(-\lambda')}$.
Its explicit expression, assuming factorization, is given by
\begin{equation}
A_{LL}d\sigma \!\!=\!\!
\sum_{ij}{1\over1+\delta_{ij}}\int\!\!
dx_adx_b\biggl[\Delta f^{(a)}_i\bigl(x_a,Q^2\bigl) \Delta
f^{(b)}_j\bigl(x_b,Q^2\bigl) \hat {a}_{LL}^{ij}
d\hat{\sigma}^{ij}+(i\leftrightarrow j)\biggl] ,
\end{equation}
where d$\sigma$ is given by eq.(2) and $\hat {a}_{LL}^{ij}$ denotes the 
corresponding subprocess double asymmetry for initial partons $i$ and $j$. The
$\Delta f$'s  are defined as
\begin{equation} 
\Delta f(x,Q^2)\; =\; f_+(x,Q^2) \, -\, f_-(x,Q^2)
\end{equation}
where $f_{\pm}$ are the parton distributions in a polarized hadron with 
helicity either parallel (+) or antiparallel (-) to the parent hadron helicity.
Recall that the unpolarized distributions are $f\, = \, f_+ \, + \,f_-$ and
$\Delta f$ measures how much the parton $f$ "remembers" the parent hadron
helicity.
If the subprocess involves parity violating interactions, one can consider 
another interesting observable which requires only {\it {one}} initial hadron
polarized, that is the {\it {single}} helicity asymmetry $A_L$, defined as
\begin{equation}
A_{L} = {d\sigma_{a(-)}-d\sigma_{a(+)}\over 
d\sigma_{a(-)}+d\sigma_{a(+)}}\, . 
\end{equation} 
In addition, if both $a$ and $b$ are polarized one can also have two
{\it {double}} helicity parity violating asymmetries defined as
\begin{equation}
A_{LL}^{PV} ={d\sigma_{a(-)b(-)}-d\sigma_{a(+)b(+)}\over 
d\sigma_{a(-)b(-)}+d\sigma_{a(+)b(+)}}
\;\; \hbox{and}\;\; 
\bar A_{LL}^{PV} ={d\sigma_{a(-)b(+)}-d\sigma_{a(+)b(-)}\over 
d\sigma_{a(-)b(+)}+d\sigma_{a(+)b(-)}} \; ,
\end{equation}
which can be simply related to $A_L$ \cite{BS93}.
Several sets of polarized parton densities $\Delta f(x,Q^2)$ ($f=q, \bar q, G$) have been proposed in the recent literature [5-10]. Using some of these parametrizations to calculate helicity asymmetries for various processes, we will show how the {\it RHIC-BNL} spin programme will be able, in particular, to pin down the polarized helicity distributions which remain badly constrained by polarized {\it DIS} experiments. 
                                                                      \section{How to pin down $\Delta G(x,Q^2)$ ?}
The cross section for direct photon production on $pp$ collisions at high
$p_T$ is considered as one of the cleanest probe of the unpolarized gluon
distribution $G(x,Q^2)$. This is partly due to the fact that the photon 
originates in the hard scattering subprocess and is detected without
undergoing fragmentation. Moreover in $pp$ collisions the quark-gluon Compton
subprocess
$qG \rightarrow q\gamma$ dominates largely and the quark-antiquark 
annihilation subprocess
$q{\bar q} \rightarrow G \gamma$ can be neglected. Consequently the double 
helicity asymmetry
$A_{LL}^{\gamma}$ (see eq.(4)), which involves in this case only one 
subprocess, becomes particularly simple to calculate and is expected to be
strongly sensitive to the sign and magnitude of $\Delta G(x,Q^2)$. For the
Compton subprocess, 
$\hat a_{LL}$ whose expression at the lowest order is given in 
ref.\cite{BRST}, is always positive and such that 
$\hat a_{LL}({\hat \theta}_{cm} = 90^{\circ})\, =\, 3/5$ where ${\hat 
\theta}_{cm}$ is the center of mass (c.m.) angle in the subprocess. 

In Fig. 1 we show a complete comparison of the results we obtained for
the $p_T^{\gamma}$ distribution of $A_{LL}^{\gamma} $, at pseudo rapidity $\eta=0$, using the 
different sets of polarized parton distributions we have mentioned in the
previous section, with a leading order $Q^2$ evolution. Actually we find that
the smallest predictions are obtained from the sets ref.\cite{GS96} and
ref.\cite{GRSV} which have the smallest
$\Delta G(x)/G(x)$. The predictions differ substantially at large $p_T$, 
which corresponds to the region, say around $x=0.4$ or so, where the
distributions
$\Delta G(x)$ have rather different shapes. We have also indicated the 
expected statistical errors based on an integrated luminosity $L = 800$
pb$^{-1}$ at
$\sqrt s = 500\,$ GeV, for three months running time. We have evaluated the 
event rates in the pseudo-rapidity gap $-1.5 < \eta < 1.5$, assuming a
detector efficiency of 100\% and for a $p_T^{\gamma}$ acceptance $\Delta
p_T^{\gamma} = 5\,$ GeV/c. We see that up to $p_T^{\gamma} = 50\,$ GeV/c or
so, $A_{LL}^{\gamma}$ will be  determined with an error less than 5\% which
therefore will allow to distinguish between these different possible $\Delta
G(x)$. For very large $p_T^{\gamma}$, the event rate drops too much to provide
any sensitivity in the determination of $\Delta G(x)$. The pseudo-rapidity distribution  of $A_{LL}^{\gamma}(\eta)$ at a fixed
$p_T^{\gamma}$ value has been also calculated in ref.(\cite{SV}) and it shows a systematic increase of $A_{LL}^{\gamma}$ for higher $\eta$.

\begin{figure}[ht]
\vspace{-1.5cm}
    \centerline{\psfig{figure=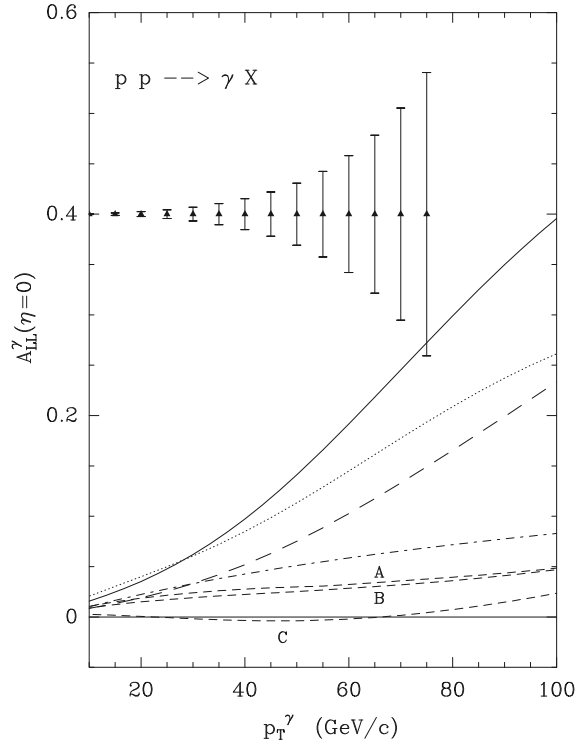,width=9cm,height=13cm}}
\vspace{-1.cm}
    \caption[]{The double helicity asymmetry $A_{LL}^\gamma(\eta=0)$
at $\eta = 0$ versus $p_T^\gamma$ for $\sqrt s = 500 GeV $ calculated with different parton densities (Solid and
large dashed curves \cite{BS95}, dotted curve \cite{BouBuc},
dashed-dotted curve \cite{GRSV}, small dashed curves
\cite{GS96}).(Taken from ref.\cite{SV})}
    \label{fig.1} 
\end{figure}

\begin{figure}[ht]
\vspace{-2.cm}
    \centerline{\psfig{figure=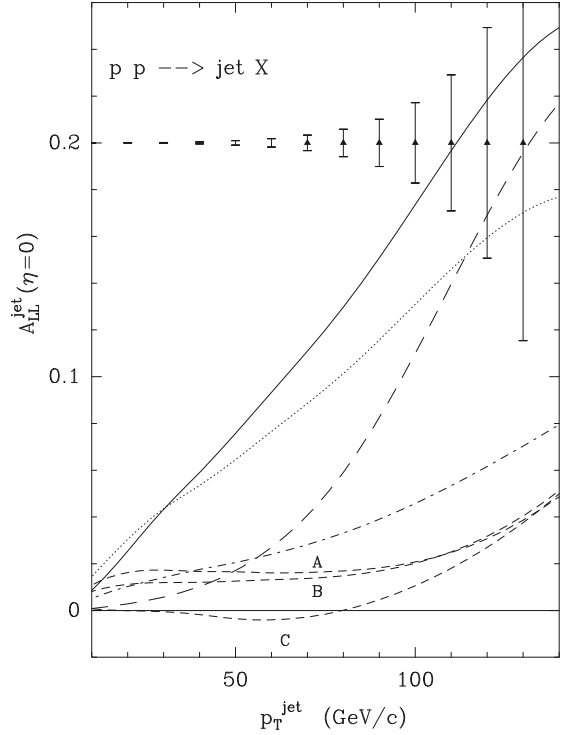,width=9cm,height=13cm}}
\vspace{-1.cm}
    \caption[]{The double helicity asymmetry $A_{LL}^{jet}(\eta=0)$
at $\eta = 0$ versus $p_T^{jet}$ for 
$\sqrt s = 500 GeV $ calculated with different parton
densities (Curve labels as in Fig.1).(Taken from ref.\cite{SV}) }
    \label{fig.2} 
\end{figure}

Inclusive jet production is also a physics area where one can learn a lot 
about parton densities and, considering the vast amount of unpolarized
existing data, it has been regarded as an important QCD testing ground. Event
rates are much larger than for prompt photon production, but there is a
drawback because many subprocesses are involved, unlike in the previous case.
In principle one should take into account gluon-gluon ($GG$), gluon-quark
($Gq$) and quark-quark ($qq$) scatterings. Although these subprocesses cross
sections are not so much different, after convolution with the appropriate
parton densities (see eq.(2)), they lead to very distinct contributions to the
hadronic spin-average cross section.
Here we will restrict ourselves
to the double helicity asymmetry
$A_{LL}^{jet}$ for single jet production and in order to clarify the 
interpretation of our results below, let us recall some simple dynamical
features. In the very low $p_T^{jet}$ region, say $p_T^{jet} \sim 10$ GeV/c or
so, $GG$ scattering dominates by far, but its contribution drops down very
rapidly with increasing $p_T^{jet}$. In the medium $p_T^{jet}$ range, say 20
GeV/c $< p_T^{jet} < 80$ GeV/c or so, 
$Gq$ scattering dominates and then decreases for large $p_T^{jet}$, to be 
overcome by
$qq$ scattering. Of course these are rough qualitative considerations and 
accurate numerical estimates for the relative fractions of these different
contributions depend strongly on the parton densities one uses.Let us look at $A_{LL}^{jet}(\eta=0)$ and, from the above discussion, we 
see  that in the medium $p_T^{jet}$ range where $Gq$ scattering dominates,
$A_{LL}^{jet}(\eta=0)$ should have a trend similar to
$A_{LL}^{\gamma}(\eta=0)$, with perhaps a magnitude reduced by a factor two,
since about half of the jet cross section is due to $GG$ and
$qq$ scatterings. This is what we see approximately in Fig.2, where we 
present the numerical results for $A_{LL}^{jet}(\eta=0)$ at ${\sqrt s} = 500$
GeV, which should be compared to Fig.1. We have also indicated the
statistical errors which are extremely small in this case, because of the huge
event rates.

\section{How to pin down $\Delta q(x,Q^2)$ and $\Delta {\bar q}(x,Q^2)$ ?}

Let us now consider, for the reaction $pp \rightarrow W^{\pm}X$, the parity 
violating single helicity asymmetry $A_L$ defined in eq.(6). In the Standard
Model, the $W$ gauge boson is a purely left-handed object and this asymmetry
reads simply for $W^{\pm}$ production 
\begin{equation}
A_L^{W^+}(y)\, =\, \frac{\Delta u(x_a,M^2_W)\, {\bar d}(x_b,M^2_W)\, -\,
(u \leftrightarrow {\bar d})}{u(x_a,M^2_W)\, {\bar d}(x_b,M^2_W)\, +\,
(u \leftrightarrow {\bar d})}\; ,
\end{equation}
assuming the proton $a$ is polarized. Here we have $x_a = {\sqrt \tau}e^y$, 
$x_b = {\sqrt \tau}e^{-y}$ and $\tau = M_W^2/s$. For $W^{-}$ production the 
quark flavors are interchanged ($u \leftrightarrow d$). The calculation of
these asymmetries is therefore very simple and the results are presented in 
Figs.3,4 at ${\sqrt s} = 500$ GeV, for different sets of distributions.
As first noticed in ref.\cite{BS93}, the general trend of $A_L$ can be easily
understood as follows : at $y=0$ one has
\begin{equation}
A_L^{W^+}\, =\, \frac{1}{2}\left( \frac{\Delta u}{u}\, -\, 
\frac{\Delta \bar d}{\bar d} 
\right)\;\;\;\;\;\;\; \hbox{and} \;\;\;\;\;\;\; A_L^{W^{-}}\, =\,
\frac{1}{2}\left( \frac{\Delta d}{d}\, -\, \frac{\Delta \bar u}{\bar u} 
\right)\; ,
\end{equation}
evaluated at $x=M_W/{\sqrt s}=0.164$, for $y=-1$ one has
\begin{equation}
A_L^{W^{+}}\, \sim\,  -\, \frac{\Delta \bar d}{\bar d} 
\;\;\;\;\;\;\; \hbox{and} \;\;\;\;\;\;\; A_L^{W^{-}}\, \sim\,
 -\, \frac{\Delta \bar u}{\bar u} \; ,
\end{equation}
evaluated at $x=0.059$ and for $y=+1$ one has
\begin{equation}
A_L^{W^{+}}\, \sim\, \frac{\Delta u}{u}\, \;\;\;\;\;\;\; \hbox{and} 
\;\;\;\;\;\;\; A_L^{W^{-}}\, \sim \,
\frac{\Delta d}{d}\; ,
\end{equation}
evaluated at $x=0.435$.
 
\begin{figure}[ht]
\vspace{-2.cm}
    \centerline{\psfig{figure=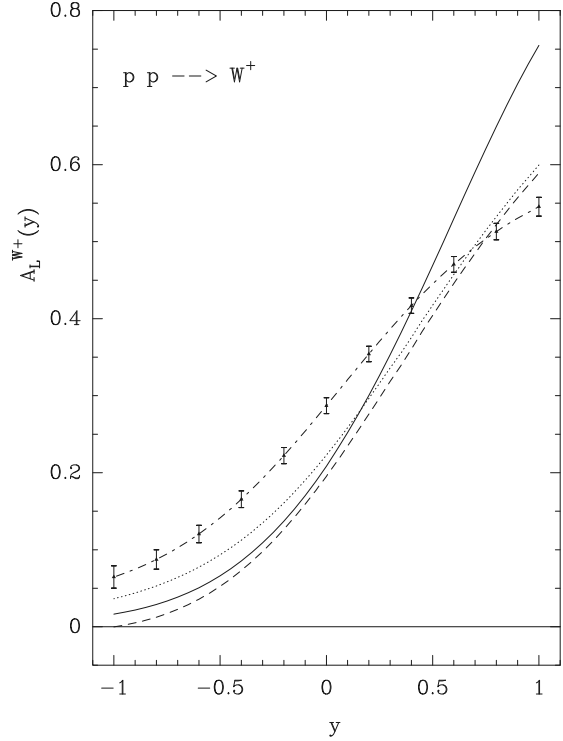,width=9cm,height=13cm}}
\vspace{-1.cm}
    \caption[]{The parity violating asymmetry $A_L^{W^{+}}(y)$ versus y
for $\sqrt s = 500 GeV$ (Taken from ref.\cite{SV}) 
calculated with different parton densities (Curve labels as in Fig.1).}
    \label{fig.3} 
\end{figure}

Therefore these measurements will allow a fairly 
clean  flavor separation, both for quarks and antiquarks, for some interesting
ranges of $x$ values. We see in Fig.3 that $A_L^{W^+}$, which is driven by the $u$ and 
$\bar d$ polarizations, leads to similar predictions for all cases. This  is
mainly due to our knowledge of $\Delta u/u$, except for $x\geq 0.3$ where it
comes out to be larger for the set proposed in ref.\cite{BS95}. In Fig.4, for 
$A_L^{W^-}$ which
is sensitive to the
$d$ and $\bar u$ polarizations, we see that the various predictions lead to 
the same general trend, with some differences in magnitude due to a large
uncertainty in the determination of $\Delta d/d$. Also in ref.\cite{BS95}, one 
has
assumed a larger negative
$\Delta {\bar u} /\bar u$, which is reflected in the behaviour near $y=-1$. The 
statistical errors have been calculated with a rapidity resolution $\Delta y =
0.2$ and taking into account only the events from the leptonic decay modes.
They are smaller for $W^+$ production which has larger event rates.
 
\begin{figure}[ht]
\vspace{-2.cm}
    \centerline{\psfig{figure=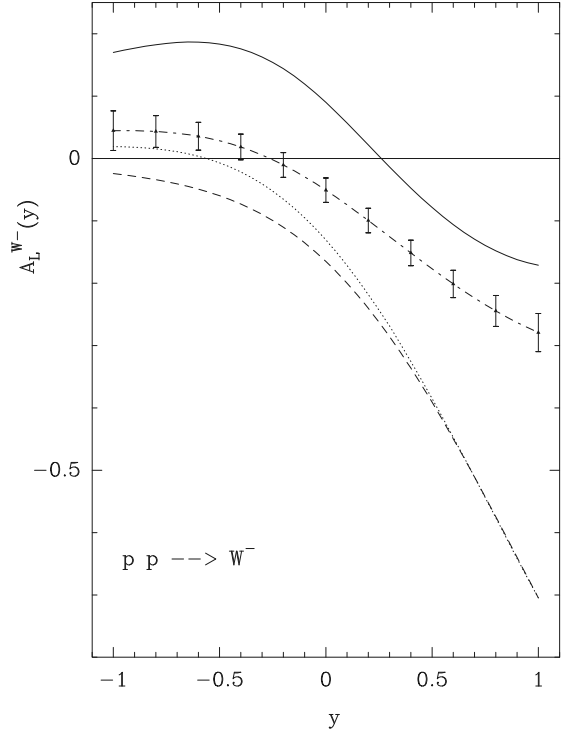,width=9cm,height=13cm}}
\vspace{-1.cm}
    \caption[]{Same as Fig.3 for $W^-$ production.}
    \label{fig.4} 
\end{figure}

\section{How to measure $h_1^{q,\bar q}(x,Q^2)$ ?}
The existence of $h_1^q(x)$ for quarks ($h_1^{\bar q}$ for antiquarks) was first detected in a systematic study
of the Drell-Yan process with polarized beams \cite{RS} and some of
its relevant properties were discussed later in various
papers \cite{AM,CPR,JJ}. Pretty much like $q(x)$ and $\Delta q(x)$, the transversity distributions
$h_1^{q,\bar q}(x)$ are of fundamental importance for our
understanding of the nucleon structure and they are all leading-twist
distributions. Due to scaling violations, these quark distributions depend
also on the scale $Q$ and their $Q^2$-behavior is predicted by the {\it QCD}
evolution equations. They are different in the three cases but, due to lake of time, we will not discuss here this important question\cite{BST,MSSV}. We recall that $h^q_1(x,Q^2)$ (or $h^{\bar q}_1(x,Q^2)$) are not simply accessible in
{\it DIS} because they are in fact chiral-odd distributions, contrarely to
$q(x,Q^2)$ and $\Delta q(x,Q^2)$ which are chiral-even~\cite{JJ}. However they can be extracted from polarized Drell-Yan processes with two transversely
polarized proton beams. For lepton pair production $pp\to\ell^+\ell^- X$
$(\ell = e, \mu)$ mediated by a virtual photon $\gamma^{\star}$, the double
transverse-spin asymmetry $A^{\gamma^{\star}}_{TT}$ defined, similarly to eq.(3), as $A_{TT}=d \delta \sigma / d\sigma$, reads

\begin{equation}
A^{\gamma^{\star}}_{TT} = \widehat a_{TT} \frac{\sum_{q}^{} e^2_q h^q_1
(x_a,M^2) h^{\bar q}_1 (x_b, M^2) + (a \leftrightarrow b)}{\sum_{q}^{} e^2_q 
q (x_a,M^2) \bar q (x_b, M^2) + (a \leftrightarrow b)},
\end{equation}
where $\widehat a_{TT}$ is the partonic asymmetry calculable in perturbative
{\it QCD} and $M$ is the dilepton mass. The rapidity $y$
of the dilepton is $y=x_a-x_b$, and for $y=0$ one has $x_a=x_b=M/\sqrt{s}$. 
 Note
that this is a leading-order expression, which can be used to get a first
estimate of
$A^{\gamma\star}_{TT}$ from different theoretical results for $h^q_1$ and
$h^{\bar q}_1$. If the lepton pair is mediated by a $Z$ gauge boson, one 
has a similar expression for $A^Z_{TT}$ \cite{BS94},namely
\begin{equation}
A^{Z}_{TT} = \frac{\sum_{q}^{} (b^2_q -a^2_q) h^q_1
(x_a,M^2_Z) h^{\bar q}_1 (x_b, M^2_Z)+ (a \leftrightarrow b)}{\sum_{q}^{} 
(b^2_q +a^2_q)q (x_a,M^2_Z) \bar q (x_b, M^2_Z)+ (a \leftrightarrow b)},
\end{equation}
where $a_q$ and $b_q$ are the vector and axial couplings of the flavor $q$ to 
the $Z$. However in the case of
$W^{\pm}$ production one expects $A^W_{TT}=0$, because the $W$ gauge boson 
is a pure left-handed object ({\it i.e.}, $a_q=b_q$), which does not allow a
left-right interference effect associated to the existence of $h^{q,\bar
q}_1$ \cite{BS94}. We show in Fig.5 some predictions, at leading and next
to leading orders, for $A_{TT}$, together with some expected statistical errors, where one sees a characteristic effect in the $Z$ mass region. The asymmetry is only a few percents, due to the smallness of
$h_1^{\bar q}$, but this is a real challenge for the experiment.

\begin{figure}[ht]
    \centerline{\psfig{figure=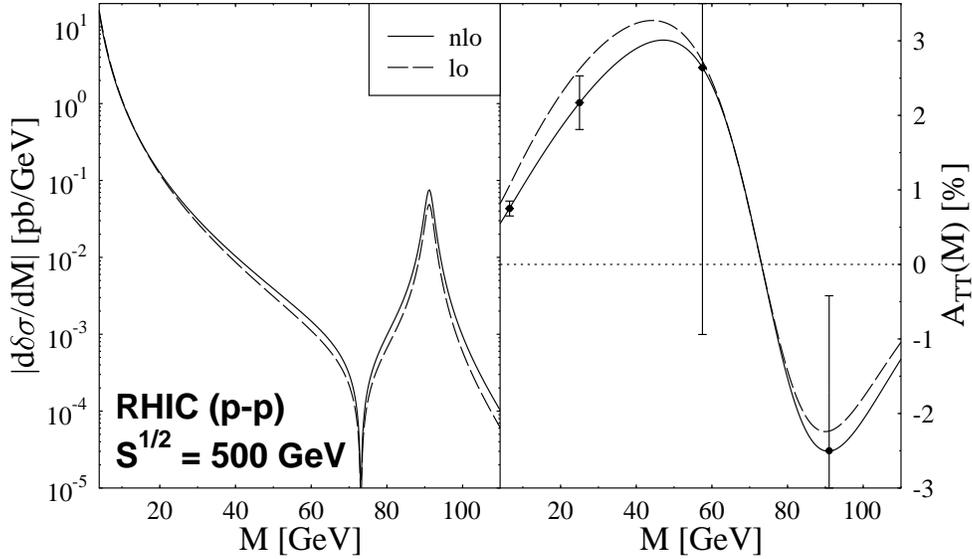,width=13.cm,height=8.5cm}}
    \caption[]{NLO and LO maximal polarized Drell-Yan cross section ($|d\delta \sigma|$) and asymmetry($A_{TT}$) for {\it RHIC} at $\sqrt s =500GeV$ assuming the maximum luminosity (Taken from ref.\cite{MSSV}).}
    \label{fig.5} 
\end{figure}

\section{How to uncover new parity violation effects ?}
Let us consider again one-jet inclusive production. As discussed in section 3, the cross section is dominated by the pure {\it QCD} subprocesses $GG$, $Gq$ and
$qq$ scatterings, but the existence of the electroweak ({\it EW}) interaction,
via the effects of the $W^{\pm},Z$ gauge bosons, adds to it, a small contribution. Consequently, the parity violating asymmetry $A_{LL}^{PV}$ defined in eq.(7) and resulting from the {\it QCD-EW} interference is non-zero, as shown
in Fig.6 (see curve {\it SM}) and has a small structure near $E_T=M_{W,Z}/2$.
\begin{figure}[ht]
\vspace{-2.cm}
\centerline{\psfig{figure=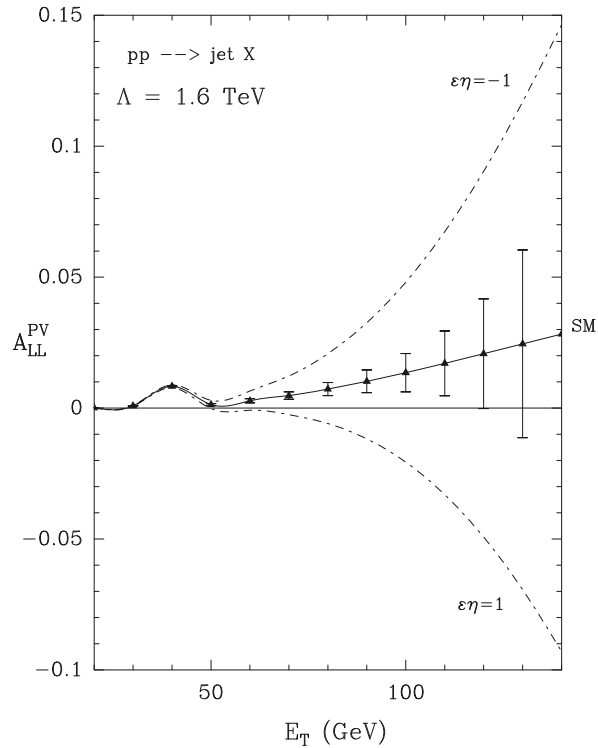,width=9.0cm,height=13.0cm}}
\vspace{-1.cm}
    \caption[]{$A_{LL}^{PV}$ versus $E_T$ at {\it RHIC}  for $\sqrt s=500GeV$, for $\Lambda=1.6TeV$ (Dashed curve corresponds to $\epsilon\eta=-1$, dot-dashed curve to $\epsilon\eta=+1$ and solid curve, {\it SM}, is the pure {\it QCD-EW} interference) (Taken from ref.\cite{TV}).}
    \label{fig.6} 
\end{figure}
Now if we introduce a new contact interaction, belonging purely to the quark sector and normalized to a certain compositeness scale $\Lambda$, under the form
\begin{equation}
{\cal L}_{qqqq} = \epsilon \, {g^2\over {8 \Lambda^2}} 
\, \bar \Psi \gamma_\mu (1 - \eta \gamma_5) \Psi . \bar \Psi
\gamma^\mu (1 - \eta \gamma_5) \Psi~,
\end{equation}
where $\Psi$ is a quark doublet and $\epsilon=\pm 1$. If parity is not conserved
$\eta=\pm 1$ and we show in Fig.6, how the {\it SM} prediction will be affected by such a new interaction assuming $\Lambda=1.6GeV$. As expected, the 
errors are large in the high $E_T$ region, but if the observation of such a signal is confirmed, it will be extremely important.

{\it It it my pleasure to thank Marek Je$\dot{\rm {\it z}}$abek for the invitation and all the other organizers, for setting up this excellent Conference in such a pleasant and stimulating atmosphere}.

\end{document}